\nonstopmode

\newcommand{\cfbr}{${\rm CF}_{3}{\rm Br}$ }
\newcommand{\cfbrn}{${\rm CF}_{3}{\rm Br}$}
\newcommand{\U}[1]{\,{\rm{#1}}}
\newcommand{\eV}{\U{eV}}

\documentclass[prl,aps,superscriptaddress,twocolumn,nopreprintnumbers]{revtex4}
\usepackage{graphicx,amssymb,bm,amsmath}
\usepackage[nativepdf,bookmarks,bookmarksopen,bookmarksnumbered,raiselinks,%
pdftitle={Using strong electromagnetic fields to control x-ray processes},%
pdfauthor={Linda Young, Christian Buth, Robert W. Dunford, Phay J. Ho, %
Elliot P. Kanter, Bertold Krässig, Emily R. Peterson, Nina Rohringer, %
Robin Santra, Stephen H. Southworth},%
pdfsubject={Atomic physics, chemical physics},%
pdfkeywords={quantum optics, ab initio, laser dressing, x rays, %
photoabsorption, cross section, polarization dependence, nonperturbative, %
laser-aligned molecules, neon, krypton, electromagnetically induced %
transparency, alignment of atomic orbitals, ionization, diffraction, CF3Br, %
bromotrifluoromethane, EIT, pulse shaping}]{hyperref}

\begin{document}
\title{Using strong electromagnetic fields to control x-ray processes}
\author{Linda Young}
\affiliation{Argonne National Laboratory, Argonne, Illinois~60439, USA}
\author{Christian Buth}
\thanks{Present address: Department of Physics and Astronomy, Louisiana State
University, Baton Rouge, Louisiana~70803, USA}
\affiliation{Argonne National Laboratory, Argonne, Illinois~60439, USA}
\author{Robert W.~Dunford}
\affiliation{Argonne National Laboratory, Argonne, Illinois~60439, USA}
\author{Phay J.~Ho}
\affiliation{Argonne National Laboratory, Argonne, Illinois~60439, USA}
\author{Elliot P.~Kanter}
\affiliation{Argonne National Laboratory, Argonne, Illinois~60439, USA}
\author{Bertold Kr\"assig}
\affiliation{Argonne National Laboratory, Argonne, Illinois~60439, USA}
\author{Emily R.~Peterson}
\affiliation{Argonne National Laboratory, Argonne, Illinois~60439, USA}
\author{Nina Rohringer}
\thanks{Present address: Lawrence Livermore National Laboratory,
Livermore, California~94551, USA}
\affiliation{Argonne National Laboratory, Argonne, Illinois~60439, USA}
\author{Robin Santra}
\affiliation{Argonne National Laboratory, Argonne, Illinois~60439, USA}
\affiliation{Department of Physics, University of Chicago, Chicago,
Illinois~60637, USA}
\author{Stephen H.~Southworth}
\affiliation{Argonne National Laboratory, Argonne, Illinois~60439, USA}
\date{September 20, 2008}

\begin{abstract}
Exploration of a new ultrafast-ultrasmall frontier in atomic and molecular
physics has begun.
Not only is is possible to control outer-shell electron dynamics with
intense ultrafast optical lasers, but now control of inner-shell processes has
become possible by combining intense infrared/optical lasers with tunable
sources of x-ray radiation.
This marriage of strong-field laser and x-ray physics has
led to the discovery of methods to control reversibly resonant x-ray
absorption in atoms and molecules on ultrafast timescales.
Using a strong optical dressing field, resonant x-ray absorption in atoms
can be markedly suppressed, yielding an example of electromagnetically
induced transparency for x~rays.
Resonant x-ray absorption can also be controlled in molecules using
strong non-resonant, polarized laser fields to align the framework of
a molecule, and therefore its unoccupied molecular orbitals to which
resonant absorption occurs.
At higher laser intensities, ultrafast field ionization produces an
irreversible change in x-ray absorption.
Finally, the advent of x-ray free electron lasers enables
first exploration of non-linear x-ray processes.
\end{abstract}

\maketitle

\section{Introduction}

\begin{figure}[tb]
  \centerline{\includegraphics[clip,width=\hsize]{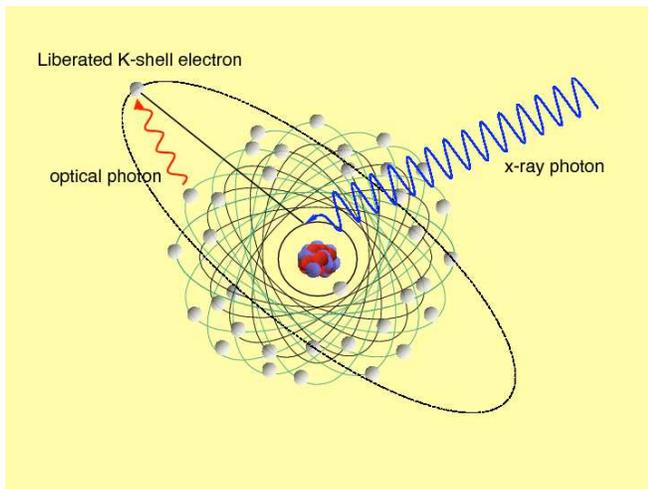}}
  \caption{(Color online) X-ray absorption promotes a $1s$~electron to an
           empty outer shell orbital.
           Optical radiation induces transitions between outer
           shell orbitals.
           Inner and outer shells are coupled through resonant x-ray
           absorption.}
  \label{fig1}
\end{figure}

Control of x-ray processes using intense optical lasers represents an emerging
scientific frontier---one which combines x-ray physics with strong-field laser
control~\cite{aamop08}.
While the past decade has produced many examples where
intense lasers at optical wavelengths are used to control molecular
motions~\cite{Rice00,Shapiro03,Rabitz00,Stap03RMP}, extension to the control
of intraatomic inner-shell processes is quite
new~\cite{aamop08,Buth07PRA,Buth07PRL,Peterson08APL}.
At first glance, it is an unusual concept to control x-ray processes using
an optical or infrared radiation field since x-rays interact predominantly
with inner-shell electrons, whereas longer wavelength radiation interacts
with outer shell electrons.
However, the inner and outer shells of atoms are coupled through resonant
x-ray absorption, e.g., promotion of a $K$-shell electron to an empty outer
shell orbital, as shown in Fig.~\ref{fig1}.
Because outer shell electronic structure can be perturbed
(dressed) by an optical radiation field, one can exert control over resonant
x-ray absorption using optical lasers.
Reversible control is possible when the applied dressing field is gentle
enough to significantly perturb outer-shell electronic
structure, but is not intense enough to destroy (ionize) the atom.

Let us consider the optical field amplitude necessary to achieve this control,
i.e., to induce outer-shell transitions at a rate comparable to inner-shell
processes.
If we take the simplest case, shown in Fig.~\ref{fig1}, where absorption of an
x-ray photon leads to the ejection of a $K$-shell electron, a $1s^{-1}$ hole is
created.
The resulting atom, containing a $1s^{-1}$~hole, is unstable and decays
via both radiative and non-radiative (Auger) channels~\cite{Krause79}.
These inner-shell decay rates increase with atomic number;
at $Z=10$ the lifetime of the $1s^{-1}$ hole state is $2.4 \U{fs}$
(corresponding to a $0.27 \eV$ level width via the uncertainty principle).
In order to compete with the rapid inner-shell decay,
electromagnetic transitions in the outer shell must be induced at a comparable
rate.
The transition rate in a driven two-level system is given by the Rabi
flopping frequency, $\Omega_{12}=\mu_{12}E/\hbar$ where $\mu_{12} =
\langle1|ez|2\rangle$ is the transition dipole matrix element between
levels~1 and 2, $E$ is the electric field amplitude.  
In practical units, $E$ is related to the
laser intensity via $I$[W/cm$^2$]$=(1/2Z_0)E^2$ [V/cm], where $Z_0 =
\sqrt{\mu_0\epsilon_0} = 377$ V/A is the vacuum impedance.
To estimate the laser intensity required to drive Rabi oscillations at a
rate comparable to inner shell decay we use atomic units which are related
to the hydrogen atom: charge, $e=1$;
length = Bohr radius = $a_0$;
velocity = Bohr velocity $v_0= \alpha c$;
electric field = field at the Bohr radius $e/a_0^2\sim 51$ V/{\AA};
electric dipole moment $= ea_0$, time $t_0 = a_0/v_0 \sim 0.024 \U{fs}$.
For the hydrogen $1s\rightarrow 2p_{1/2}$~transition, $\mu_{12} = 1.05 ea_0$.
Combining this with an electric field amplitude $E=1$ atomic unit, the Rabi
flopping frequency will be $\Omega_{12}=1/t_0 = 1/0.024 \U{fs}$.
The laser-induced rate is $16\times$ that of the decay rate of the $1s$~hole
state in neon!
($\Omega_{12}$ are given in angular frequency units)
Therefore, if the electric field amplitude of the optical laser can be
1~atomic unit, or equivalently a laser intensity of $3.5 \times 10^{16}
\U{W/cm^2}$, outer-shell transition rates can
exceed inner-shell decay rates.

This intensity can now be achieved with routinely available modern
lasers~\cite{Krausz00,Mourou06RMP}.
The key development of chirped pulse amplification~\cite{Strick85OL}
permits increased output energies from ultrashort lasers without damage
to amplifying media.
Focusing a standard amplified Ti:sapphire laser ($800 \U{nm}$, $3 \U{mJ}$,
$40 \U{fs}$, $1 \U{kHz}$) to $15 \U{\mu m}$ produces $I\sim 3.5 \times
10^{16} \U{W/cm^2}$.
Ti:sapphire lasers are very versatile;
the wide bandwidth allows one to stretch the pulsewidth from tens of
femtoseconds to hundreds of picoseconds, and permitting the study of
atoms and molecules exposed to a wide range of electromagnetic field
strengths.

\begin{figure}[tb]
  \centerline{\includegraphics[clip,width=\hsize,angle=0]{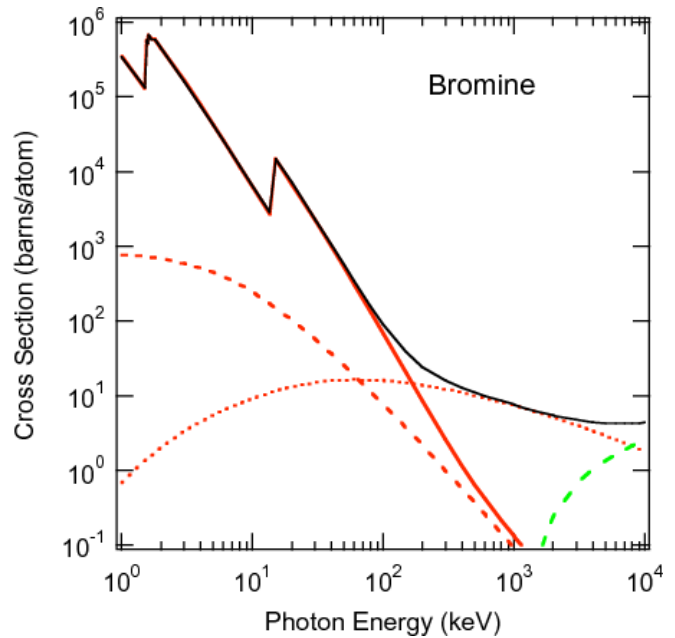}}
  \caption{(Color online) X-ray photoprocesses for bromine.
           Photoabsorption (solid line) is the dominant process;
           the $K$ and $L$~edges are visible.
           Rayleigh scattering (dashed line).
           Compton scattering (dotted line).
           Synchrotron radiation covers the range from~$10 \eV$
           to $100,000 \eV$.}
  \label{fig2}
\end{figure}

After the strong optical laser field, the next ingredient needed to study
control of ultrafast inner-shell processes is a tunable x-ray source.
Synchrotrons provide a convenient source of pulsed, tunable, polarized
radiation from $10 \eV$ to $100,000 \eV$.
This range covers inner shell edges of all elements.
In Fig.~\ref{fig2} the three dominant photoprocesses, photoabsorption,
elastic (Rayleigh) scattering and inelastic (Compton) scattering are shown
for the bromine atom.
Photoabsorption cross sections greatly exceed scattering cross sections over
energy ranges from below to far above the respective $K$~edges of each
atom.
In order to probe atoms and molecules subjected to strong pulsed optical
fields, a short x-ray pulse is helpful.
A typical x-ray pulse length is $100 \U{ps}$ at the Argonne Advanced Photon
Source.
For a $100 \U{ps}$ x-ray pulse duration, one can probe atoms and molecules
subjected to $10^{12} \U{W/cm^2}$ with millijoule laser pulse energies
focused to tens of microns.
Shorter pulse lengths, $\sim 100 \U{fs}$, at synchrotron sources are
currently available using laser slicing techniques conceived by Zholents
and Zolotorev~\cite{Zholents95PRL} and put into practice at the Advanced
Light Source~\cite{Schoenlein00Sci}, BESSY~\cite{Khan06PRL} and
the Swiss Light Source~\cite{Beaud07PRL}.

\begin{figure*}[tb]
  \centerline{\includegraphics[clip,width=\hsize]{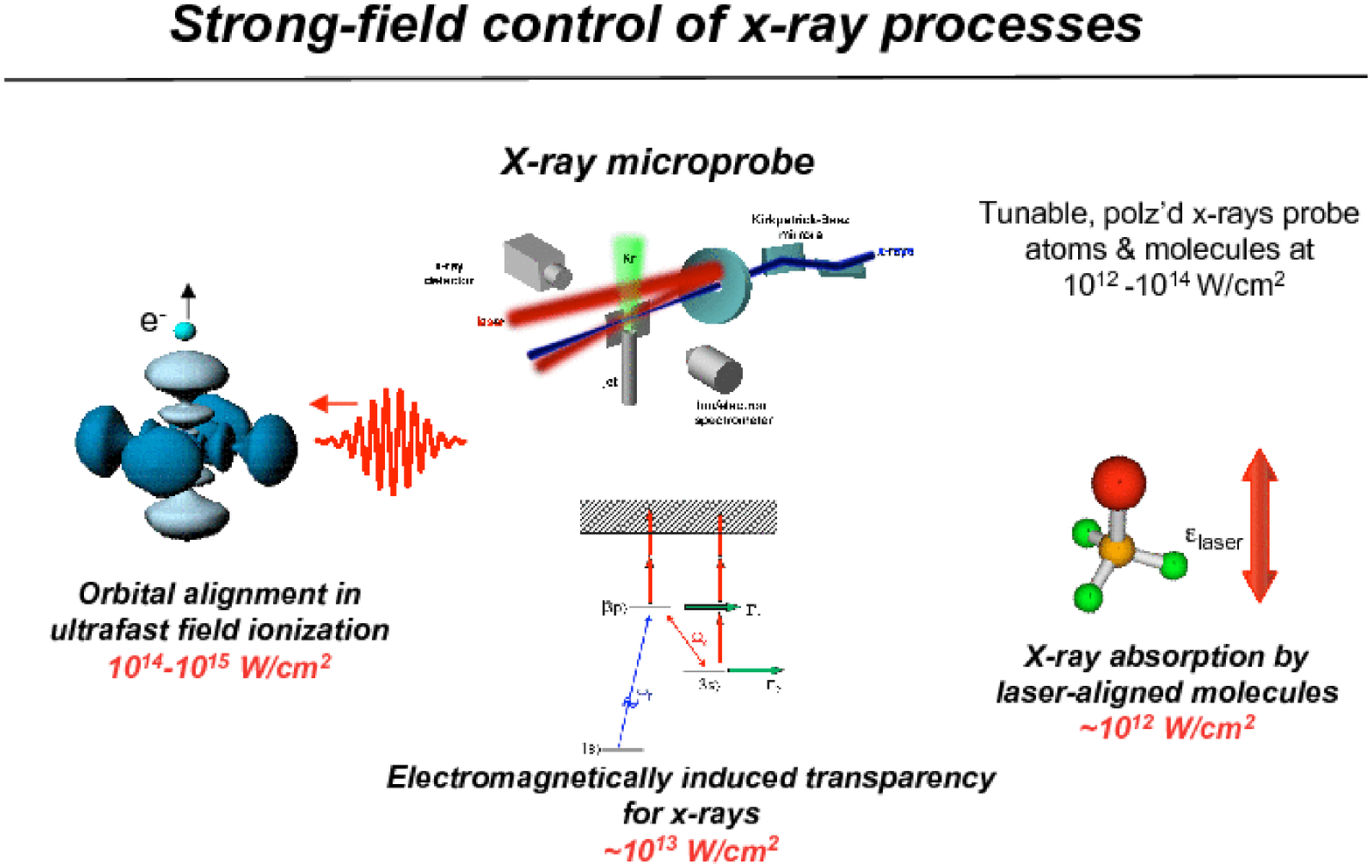}}
  \caption{(Color online) X-ray microprobe and various scenarios
           for laser control of x-ray absorption at field strengths
           ranging from $10^{12}$--$10^{15} \U{W/cm^2}$.}
  \label{microprobe}
\end{figure*}

We combine the ultrafast, ultrasmall optical and x-ray pulses using the x-ray
microprobe method~\cite{Young06PRL}.
Basically, focused laser pulses ($30 \U{\mu m}$)
are overlapped in time and space with microfocused x-ray pulses
($10 \U{\mu m}$).
Gas-phase systems are particularly suitable for illustrating the basic
principles underlying laser control of ultrafast x-ray processes.
We discuss two different scenarios for modifying resonant x-ray absorption
near an inner-shell edge:
(1)~modification of electronic structure of inner-shell-excited systems
by laser dressing at~$10^{12}$--$10^{13} \U{W/cm^2}$;
(2)~control of resonant x-ray absorption by molecules through laser-induced
spatial alignment at $10^{11}$--$10^{12} \U{W/cm^2}$.
We also discuss modifications to elastic scattering patterns from an
ensemble of laser-aligned molecules.
Another method to modify resonant x-ray absorption is through strong-field
ionization of the target particles at laser intensities in the range $10^{14}$--
$10^{15} \U{W/cm^2}$~\cite{Young06PRL,Santra06PRA,Southworth07PRA}.
Fig.~\ref{microprobe} illustrates the microprobe and these applications.

Beyond optical laser control of x-ray processes, it is also possible for strong
electromagnetic fields at x-ray wavelengths to modify characteristic x-ray
processes.
The high x-ray intensities required for these modifications will soon
be available with x-ray free electron lasers~\cite{LCLS02,Tanaka:SCSS-05,%
Altarelli:TDR-06}.

\section{Control of x-ray processes with strong optical fields}
\subsection{Electromagnetically induced transparency for x~rays: atoms}

\begin{figure*}[tb]
  \centerline{\includegraphics[width=5in,angle=0]{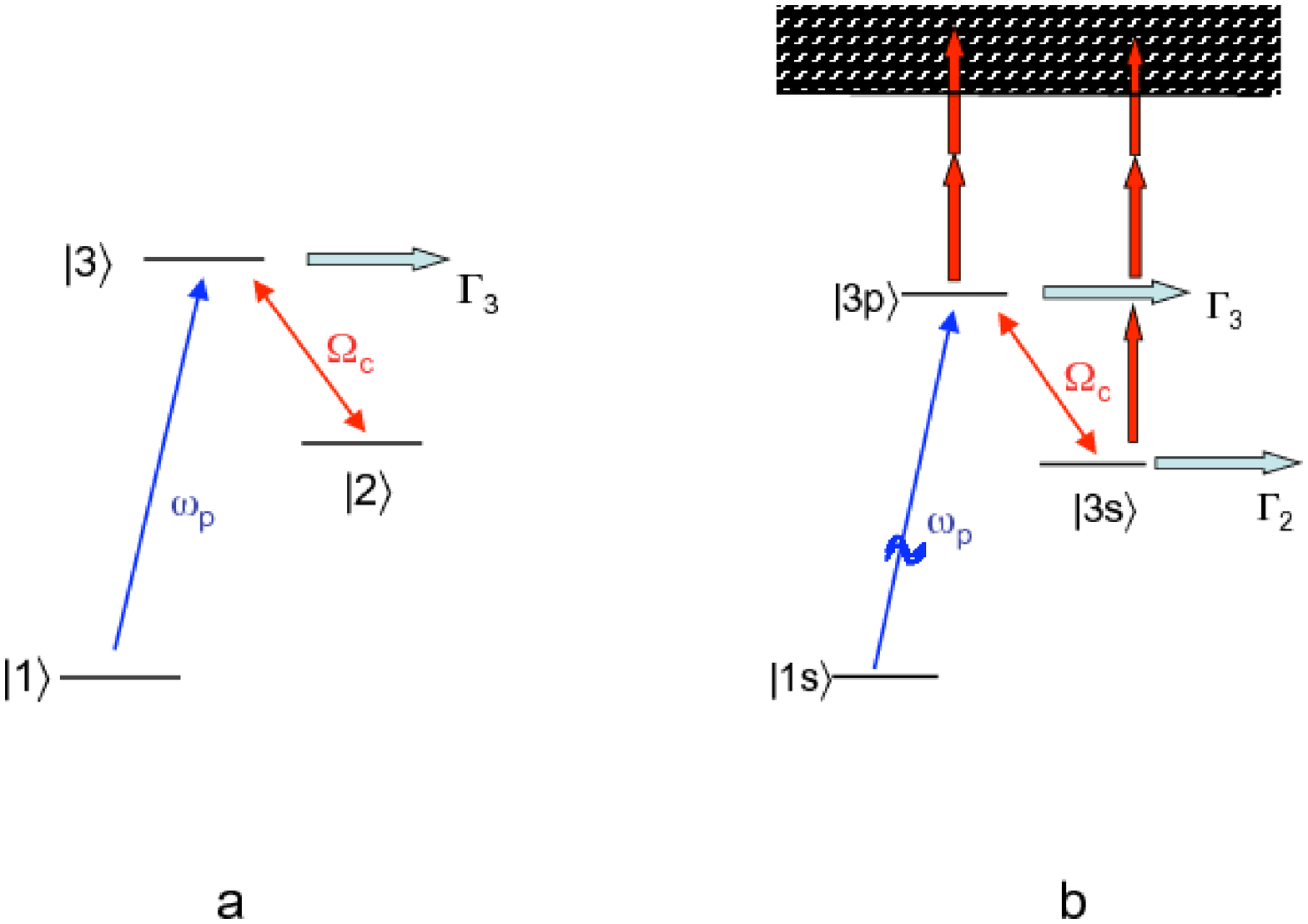}}
  \caption{(Color online) (a)~Optical EIT.
           (b)~X-ray EIT in neon.}
  \label{EIT}
\end{figure*}

In this subsection we describe laser-induced modifications to x-ray absorption
spectra in various rare gases.
Theoretical considerations are described in detail
elsewhere~\cite{Buth07PRA,Buth07PRL,Buth08proc}.
Here we focus on the analogy to electromagnetically induced transparency
in the optical regime and the application to imprinting ultrashort optical
pulse sequences and shapes onto longer x-ray pulses.

Electromagnetically induced transparency in the optical regime has been widely
studied~\cite{Harris90PRL,Boller91PRL,Harris97PT}.
In a $\Lambda$-type medium characterized by atomic
levels~$|1\rangle,|2\rangle$, and $|3\rangle$ with
energies~$E_1 < E_2 < E_3$, resonant absorption on
the $|1\rangle\rightarrow|3\rangle$~transition can be strongly suppressed
by simultaneously irradiating the medium with an intense laser that couples
the levels~$|2\rangle$ and $|3\rangle$.
This phenomena, shown in Fig.~\ref{EIT}a, is known as electromagnetically
induced transparency, EIT.

In the x-ray regime, EIT is considerably more complex.
In the optical regime, levels~$|1\rangle$ and $|2\rangle$ are stable to
electronic decay.
However, in the x-ray regime the core-excited states, $|2\rangle$ and
$|3\rangle$ are metastable.
For the three-level system in neon, depicted in Fig.~\ref{EIT}b, the
lifetime for these core-excited states is~$2.4 \U{fs}$ ($0.27 \eV$).
Thus, at an optical field strength sufficient to compete with inner-shell
decay, multiphoton transitions to the continuum can also play a role.
Inner-shell decay rates are denoted by~$\Gamma_2$ and $\Gamma_3$
and multiphoton transitions to the continuum are denoted by red block arrows.

\begin{figure}[tb]
  \centerline{\includegraphics[width=3in,angle=0,clip]{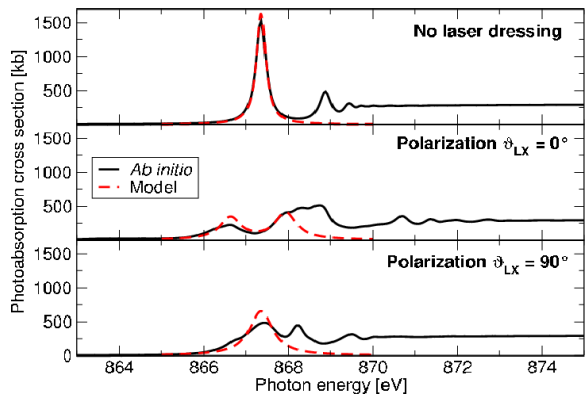}}
  \caption{(Color online) Photoabsorption near the
           $1s\rightarrow3p$~resonance in neon.
           Results from {\em ab initio} calculations and a three-level
           model are shown.
           From~\cite{Buth07PRL}.}
  \label{neonEIT}
\end{figure}

For the model system in neon, the calculated x-ray photoabsorption cross
section for~$800 \U{nm}$ laser dressing of the $1s\rightarrow 3p$~transition
in neon at $10^{13} \U{W/cm^2}$ with parallel and perpendicular laser/x-ray
polarizations along with a fit to a three-level EIT model~\cite{Buth07PRL}
is reproduced here for convenience in Fig.~\ref{neonEIT}.
At this intensity the $1s\rightarrow3p$~excitation at~$867 \eV$ is
suppressed by a factor of~13 for the configuration in which the laser and
x-ray polarizations are parallel.
As can be seen in Fig.~\ref{neonEIT}, the three level model reproduces
most features of the calculated laser-dressed x-ray photoabsorption spectrum.
In the three-level model effective linewidths of~$\Gamma_3 = 0.68 \eV$
and $\Gamma_2 = 0.54 \eV$ account for the laser-ionization broadening.
These parameters effectively reproduce both the lineshapes for the
parallel and perpendicular configurations in the vicinity of the
$1s\rightarrow3p$~resonance.

\begin{figure}[tb]
  \centerline{\includegraphics[width=3in,angle=0]{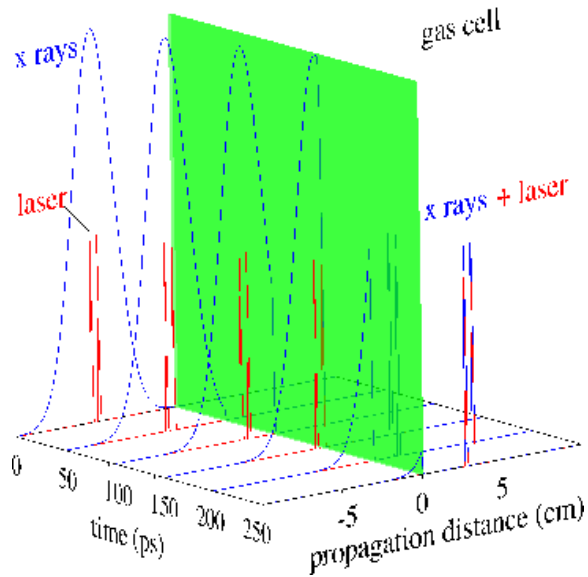}}
              \caption{(Color online) Generation of ultrafast x-ray pulses
                       using laser dressing of Ne.
                       See the text for details.}
  \label{neoncell}
\end{figure}

The ability to control x-ray absorption in Ne at the
$1s\rightarrow3p$~resonance allows one to imprint pulses shapes of the
optical dressing laser onto long x-ray pulses~\cite{Buth07PRL}.
This idea is illustrated in Fig.~\ref{neoncell}.
With a $2 \U{mm}$-long gas cell with one atmosphere of neon, the
transmission of an x-ray pulse resonant with the $1s\rightarrow3p$~transition
will be only $0.07\%$.
A typical x-ray pulse from a synchrotron source has a duration of~$100 \U{ps}$.
Such an x-ray pulse may be overlapped in time and space with one or several,
ultrashort intense laser pulses.
Those portions of the x-ray pulse that overlap with the laser are transmitted
through the gas cell.
In the case shown in Fig.~\ref{neoncell}, where the two dressing laser pulses
have a peak intensity of~$10^{13} \U{W/cm^2}$, the intensity
of the two transmitted x-ray pulses is roughly~$60\%$ of the incoming pulse.
The time delay between the two x-ray pulses can be controlled by changing
the time delay between the two laser pulses, opening a route to ultrafast
all x-ray pump-probe experiments.
With an analogous strategy, controlled shaping of short-wavelength pulses
might become a reality.
A disadvantage of this method is that it is applicable only at certain
x-ray energies;
the Ne case operates at~$867 \eV$.

It would be interesting to be able to extend these concepts to the hard x-ray
regime.
Calculations on the laser-dressed spectra near the $K$~edge in
krypton~\cite{Buth07PRA} and argon~\cite{Buth08tbp} have been done.
In argon and krypton, $K$~edges are at~$3.2 \U{keV}$ and $14.3 \U{keV}$,
respectively.
The difficulty is that the inner shell decays are significantly more rapid;
for~Ne, Ar, Kr the decay widths of core-excited $1s^{-1}np$ levels
are~$0.27$, $0.6$ and $2.7 \eV$.
The increased widths have two consequences.
First, the $1s\rightarrow np$~transitions are not fully resolved as the
level width increases, making it difficult to isolate a three-level system.
For example, the $1s^{-1}5p$, $1s^{-1}6p$,~transitions near
the krypton $K$~edge are blended such that it is smooth edge.
Second, the laser intensity needed to perturb the x-ray absorption
cross section roughly scales as the square of the level width.
Thus, when a dressing field of $10^{13} \U{W/cm^2}$ is applied, the
Kr~absorption cross section changes by less than~$20\%$.

\begin{figure}[tb]
  \centerline{\includegraphics[clip,width=\hsize]{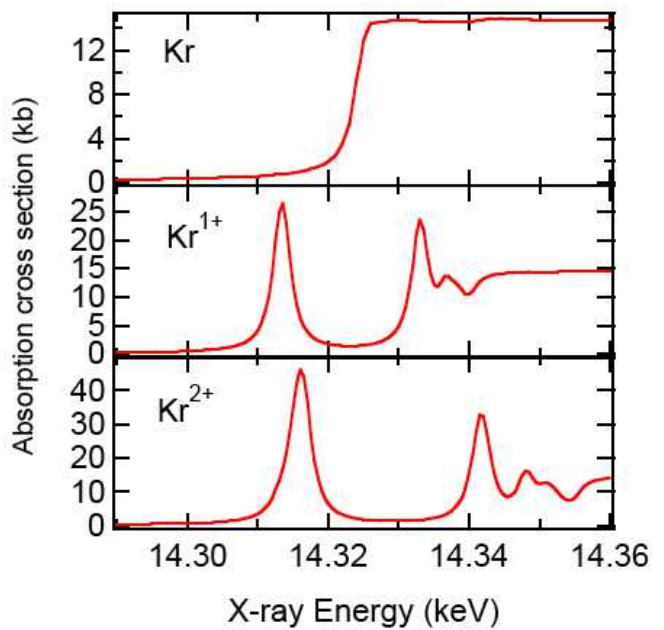}}
  \caption{(Color online) X-ray absorption spectra for Kr, Kr$^{1+}$
           and Kr$^{2+}$.}
  \label{krion}
\end{figure}

One possible way to overcome these issues is to use rare gas ions as the
EIT~medium.
The singly ionized krypton atom has a prominent, isolated resonance due
to the $4p$~hole orbital~\cite{Young06PRL,Southworth07PRA,Pan05JPB}.
X-ray absorption spectra for~Kr, Kr$^{1+}$ and Kr$^{2+}$ are shown in
Fig.~\ref{krion}.
In addition to the isolated resonance, ions can withstand higher dressing
intensities without ionization.
The critical field strength, $E_{crit}$ that one can apply to an atom
without ionizing it can be estimated by combining the laser
and Coulomb potential, $E_{crit}=E^2_{IP}/(4Ze^3)$, where $E_{IP}$ is the
ionization potential, $Z$ is the charge of the residual ion, $e$ is the
electron charge~\cite{Augst89PRL}.
For~Kr, Kr$^{1+}$ and Kr$^{2+}$ the ionization potentials are~$14.0$,
$27.9$ and $41.8 \eV$, respectively.
Thus, one should be able to apply a laser intensity~$9\times$ higher
for~Kr$^{2+}$ than neutral krypton to generate a sizable change in
absorption cross section.
Of course, generating a sufficient density of ions to realize a practical
hard x-ray pulse shaper would be a challenge.

\subsection{X-ray absorption and scattering using laser-aligned molecules}

In the presence of a strong non-resonant linearly polarized laser field,
molecules align due to the interaction of the laser electric field vector
with the anisotropic molecular polarizability~\cite{Friedrich95,Stap03RMP}.
The alignment process is of intrinsic interest and of interest in applications
to spectroscopy and photophysics, quantum control of molecular dynamics,
high-harmonic generation, chemical reactivity, liquids and solvation and
structural determinations by x-ray or electron diffraction.
For asymmetric molecules having three distinct moments of inertia,
3D~alignment has been achieved both in the presence of a laser-aligning
field (adiabatic alignment)~\cite{Larsen00} and under field-free
conditions (impulsive alignment)~\cite{Lee06}.
X-ray probes of laser-aligned molecules are of considerable current
interest~\cite{Peterson08APL} because of proposals to determine structure
of non-periodic specimens, such as individual biomolecules, by x-ray
scattering~\cite{Neutze00,Spence04} using x-ray free electron lasers.

\begin{figure}[tb]
  \centerline{\includegraphics[clip,width=\hsize]{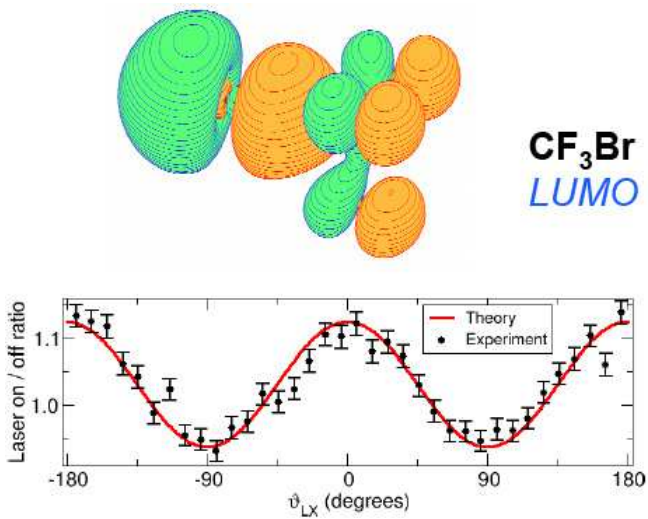}}
  \caption{The LUMO of~\cfbr is an antibonding $\sigma^*$ orbital with
           substantial Br$\,4p_z$ character, where $z$ refers to the
           C--Br~axis.
           X-ray absorption on the $1s\rightarrow\sigma^*$~resonance as a
           function of the angle between the polarization axis of the
           aligning laser and that of the x-ray probe.}
  \label{LUMOwveplt}
\end{figure}

Both x-ray absorption and scattering from a randomly oriented ensemble of gas
phase molecules are significantly modified if the molecules are aligned.
In x-ray absorption, both the near-edge and extended structure [extended x-ray
absorption fine structure~(EXAFS)] can be significantly altered.
The near-edge structure is altered when one uses resonant polarized x-ray
absorption to detect the alignment of the molecular axis relative to the
lab frame~\cite{Peterson08APL}.
Excitation of a $1s$~electron to the lowest unoccupied molecular orbital,
LUMO, often results in an isolated resonance near the {\it K}-edge of a
molecule.
The LUMO is fixed relative to the molecular frame and the transition strength
from a highly localized $K$~shell orbital to the LUMO defines the alignment
of the molecule relative to the x-ray polarization axis~\cite{Stohr}.
For the molecule that we studied, \cfbrn, the LUMO is an antibonding
$\sigma^*$~orbital with substantial Br$\,4p_z$~character, where $z$~refers
to the C--Br axis.
The molecular symmetry dictates that x-ray absorption on the Br$\,
1s\rightarrow\sigma^*$~resonance occurs only when the x-ray polarization
vector has a non-vanishing projection on the C--Br axis.
Since the laser polarization axis defines the molecular alignment
axis, one achieves control of x-ray absorption by simply rotating a waveplate.
After aligning in space, time and x-ray energy, i.e., tuning to the
$1s \rightarrow \sigma^*$~resonance at~$13.476 \U{keV}$, the control of
resonant x-ray absorption in the near-edge region can be readily
achieved~\cite{Peterson08APL,Buth08PRA}, as shown in Fig.~\ref{LUMOwveplt}.

\begin{figure}[tb]
  \centerline{\includegraphics[clip,width=3in]{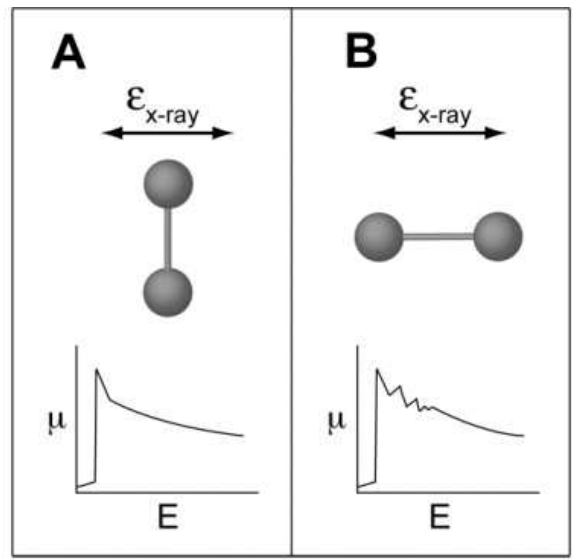}}
  \caption{Schematic of the polarization dependence of the EXAFS
           modulations in~Br$_2$.}
  \label{Br2exafs}
\end{figure}

In the EXAFS region, modifications to the polarized x-ray absorption will also
be induced by molecular alignment.
Pictorially, the effect for Br$_2$ is shown in Fig.~\ref{Br2exafs}.
When the Br$_2$ molecules are aligned parallel to the x-ray
polarization axis, the modulation of the absorption cross section due to
Br$\,1s \rightarrow \epsilon \, p$~photoelectron backscattering from the
adjacent atom is maximal, and in a perpendicular configuration minimal.
Thus, the magnitude of the EXAFS modulation is controlled by the angle of
the molecular axis relative to the x-ray polarization axis.
The equation describing the modulation around an atomic background is given
by
\begin{equation}
  \label{polexafs}
  \begin{array}{rcl}
    \displaystyle \chi(k) &=& \displaystyle -\sum_i \frac{3 \,
      \cos^2 \vartheta_i}{k \, R^2_i} \>
      F_i(k) \, e^{-2\sigma_ik^2} \, e^{-2R_i/\lambda_i(k)} \\[4ex]
    &&\displaystyle{} \times \sin(2kR_i+2\delta_l+\beta_i) \; ,
  \end{array}
\end{equation}
where $k$~is the photoelectron wavevector, $\vartheta_i$ is the angle
between the x-ray polarization and the $i^{th}$~atom, $R_i$~is the distance
between the central and $i^{th}$~atom, $\sigma_i$~is the Debye-Waller factor,
$\Lambda_i$~is the range parameter for photoelectron scattering and
$\delta_l$ and $\beta_i$~describe the outgoing photoelectron phase
shift~\cite{Stohr}.
Polarized EXAFS measurements on laser-aligned molecules will permit one
to determine changes in structure due to e.g. laser aligning fields,
though the utility of EXAFS for absolute bond distances may be limited.
Laser control of the angle between the molecular and x-ray polarization
axes will permit control of the EXAFS modulation amplitude, determination
of the atomic background and, for more complex molecules, the determination
of bond angles.

\begin{figure}[tb]
  \centerline{\includegraphics[clip,width=\hsize]{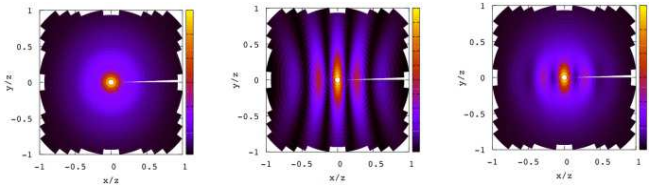}}
  \caption{Elastic scattering from isotropic, fixed-in-space and
           aligned Br$_2$ with $\langle\cos^2 \theta\rangle>(t) = 0.80$.}
  \label{br2scatt}
\end{figure}

Elastic scattering will also be modified since the contributions from the
aligned molecules can be summed linearly and a scattering pattern reflective
of the molecular structure will emerge.
A simulation of the x-ray scattering pattern from an ensemble of molecules
aligned by a laser pulse of arbitrary shape was made.
First the molecular response to the laser pulse was
calculated~\cite{Buth08PRA} and then the x-ray scattering~\cite{Ho08tbp}
from this ensemble was simulated.
In Fig.~\ref{br2scatt} elastic scattering patterns for an isotropic sample,
for clamped nuclei and for an alignment of $\langle\cos^2 \theta\rangle(t)
= 0.80$ are shown.
As the standard method for absolute molecular structure determination,
x-ray diffraction from laser-aligned molecules will then allow us to
investigate changes in molecular structure in the presence
of strong laser aligning fields with the eventual goal of a predictive
understanding using an {\em ab initio} electronic structure code such as
\textsc{dalton} to calculate molecular hyperpolarizabilities.

\section{Modifying characteristic x-ray processes with strong x-ray fields}

\begin{figure}[tb]
  \centerline{\includegraphics[clip,width=\hsize]{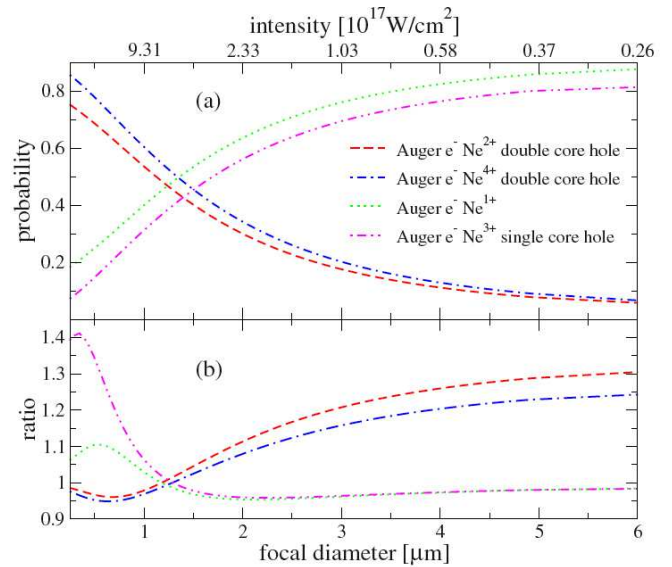}}
  \caption{(Color online) Top: Yield of~Ne[K] and Ne[KK] as a function
           of focused LCLS intensity at a photon energy of~$1050 \eV$.
           Bottom:~Ratio of Auger yields from an ensemble average of
           chaotic SASE pulses to that from an averaged pulse.
           From~\cite{Rohringer07PRA}.}
  \label{NeKK}
\end{figure}

In the previous section we have seen how it is possible to control resonant
x-ray absorption with optical fields.
With focused x-ray free electron lasers it will be possible to create
intensities such that characteristic x-ray processes are altered.
A simple estimate may be made for the $1s\rightarrow 3p$ transition
in neon.
Here the transition dipole matrix element is~$0.01 ea_0$~\cite{Rohringer08PRA}
rather than the $1.05 ea_0$ for the hydrogen $1s\rightarrow
2p_{1/2}$~transition.
Equating the Rabi frequency with the core-excited decay
rate~$(2.4 \U{fs})^{-1}$ suggests that an intensity of~$10^{18} \U{W/cm^2}$
is required to saturate this transition.
Focusing the LCLS output at~$800 \eV$ ($10^{13}$ photons / $233 \U{fs}$)
to a spot size of~$\sim 1 \U{\mu m}$ will yield this intensity.
At high intensity, the ejection of a second electron from the $1s$~shell
can be more rapid than Auger decay to create hollow neon,
Ne[KK]~\cite{Rohringer07PRA}.
The intensity dependence is shown in Fig.~\ref{NeKK}.
As one can see, hollow neon is formed preferentially at higher intensity;
indeed Ne[KK] can be the dominant species, compared to the usual
$1\%$~fraction in the weak field limit.
Because the LCLS output originates from SASE (self-amplified spontaneous
emission), the output radiation is chaotic, consisting of a random number
of intensity spikes (coherent regions) of random amplitude.
Therefore, one might expect an enhancement of two-photon processes
(such as Ne[KK] formation).
In the lower part of Fig.~\ref{NeKK}, the Auger yield from an ensemble
average of chaotic pulses is compared with the yield from an average
pulse shape.
As one can see, the chaotic enhancement for~Ne[KK] production is at most
a factor of~$1.3$ in the unsaturated regime.
This is in contrast to visible radiation where one expects a factor of~2.
The critical quantity is the ratio of the FEL coherence time to the
Auger lifetime.
The larger the ratio, the closer one will be to the maximal enhancement of~2.
Finally, we note that fully stripped neon can be produced in
the~$1 \U{\mu m}$ focus of the LCLS beam by sequential single photon
processes, providing the photon energy is greater than the
binding energy of Ne$^{9+}$~\cite{Rohringer07PRA}.

\section{Outlook}

The era in which characteristic x-ray processes can be considered invariant
is at an end.
We have demonstrated that placing atoms and molecules in strong optical
fields can significantly affect resonant absorption and elastic scattering.
Currently, we can control x-ray absorption with application of a
strong-optical field to a gaseous medium;
the control mechanism is EIT in atoms and laser-constrained rotation in
molecules.
One may be able to create an x-ray amplitude pulse shaper using these tools.
X-ray scattering from aligned molecules is not far off;
simple estimates show that a mere $10^8$ x-rays/pulse at~$1 \U{kHz}$ will
suffice as obtainable with a pink beam at the Advanced Photon
Source~\cite{Peterson08APL}.
In addition, we look forward to being able to experimentally observe
multiphoton processes in the hard x-ray regime when LCLS is
first operational in~2009.
Control of molecular alignment combined with subsequent ion/electron
imaging techniques will provide a means to disentangling
x-ray damage mechanisms in complex molecules.

\begin{acknowledgments}
We thank Dohn A.~Arms, Eric M.~Dufresne, Eric C.~Landahl, and
D.~Walko for valuable assistance during experiments at the Sector~7
beamline at the Advanced Photon Source.
C.B.~was partly supported by a Feodor Lynen Research Fellowship from the
Alexander von Humboldt Foundation.
This work and the Advanced Photon Source were supported by the Chemical
Sciences, Geosciences, and Biosciences Division of the Office of Basic Energy
Sciences,Office of Science, U.S. Department of Energy, under Contract
No.~DE-AC02-06CH11357.
\end{acknowledgments}

\end{document}